**Spin-orbit tuned metal-insulator transitions in single-crystal $Sr_2Ir_{1-x}Rh_xO_4$ ($0 \le x \le 1$)**

T. F. Qi[1], O. B. Korneta[1], L. Li[1], K. Butrouna[1], V. S. Cao[1#], Xiangang Wan[2], P. Schlottmann[3], R. K. Kaul [1] and G. Cao[1*]

[1]Center for Advanced Materials and Department of Physics and Astronomy

University of Kentucky, Lexington, KY 40506, USA

[2]Department of Physics, Nanjing University, Nanjing, P. R. China

[3]Department of Physics, Florida State University, Tallahassee, FL 32306, USA

$Sr_2IrO_4$ is a magnetic insulator driven by spin-orbit interaction (SOI) whereas the isoelectronic and isostructural $Sr_2RhO_4$ is a paramagnetic metal. The contrasting ground states have been shown to result from the critical role of the strong SOI in the iridate. Our investigation of structural, transport, magnetic and thermal properties reveals that substituting 4d $Rh^{4+}$ ($4d^5$) ions for 5d $Ir^{4+}$($5d^5$) ions in $Sr_2IrO_4$ directly reduces the SOI and rebalances the competing energies so profoundly that it generates a rich phase diagram for $Sr_2Ir_{1-x}Rh_xO_4$ featuring two major effects: **(1)** Light Rh doping ($0 \le x \le 0.16$) prompts a simultaneous and precipitous drop in both the electrical resistivity and the magnetic ordering temperature $T_C$, which is suppressed to zero at x = 0.16 from 240 K at x=0. **(2)** However, with heavier Rh doping ($0.24 < x < 0.85$ ($\pm 0.05$)) disorder scattering leads to localized states and a return to an insulating state with spin frustration and exotic magnetic behavior that only disappears near x=1. The intricacy of $Sr_2Ir_{1-x}Rh_xO_4$ is further highlighted by comparison with $Sr_2Ir_{1-x}Ru_xO_4$ where $Ru^{4+}$($4d^4$) drives a direct crossover from the insulating to metallic states.

## I. Introduction

$Sr_2IrO_4$ is an archetype for new physics primarily driven by the interplay of electron-electron and spin-orbit interactions (SOI) **[1-3]**. The relativistic SOI proportional to $Z^4$ (Z is the atomic number) is approximately 0.4 eV in the iridate (compared to ~ 20 meV in 3d materials), and splits the $t_{2g}$ bands into bands with $J_{eff}$ = 1/2 and $J_{eff}$ = 3/2, the latter having lower energy **[1-2]**. Since the $Ir^{4+}$ ($5d^5$) ions provide five 5d-electrons, four of them fill the lower $J_{eff}$ = 3/2 bands, and one electron partially fills the $J_{eff}$ = 1/2 band where the Fermi level $E_F$ resides. The $J_{eff}$ = 1/2 band is so narrow that even a reduced on-site Coulomb repulsion U (~ 0.5 eV) due to the extended nature of 5d-electron orbitals is sufficient to open a small gap (≤0.1 eV) supporting the insulating state **[1, 2]**. Most recently, an x-ray absorption spectroscopy study indicates a mixing of the $J_{eff}$ = 1/2 and $J_{eff}$ = 3/2 bands as a result of exchange interactions (~ 0.2 eV) and a tetragonal crystal electric field (CEF) (~ 0.075 eV) **[4]**. Nevertheless, the larger the SOI and the narrower the band is, the smaller U is needed for a SOI-related insulating state **[5]**, in which SOI, Coulomb interactions, tetragonal CEF and Hund's coupling $J_H$ become so comparable that they vigorously compete with each other, setting a new balance between the relevant energies that can drive new exotic states **[1-18]**.

In contrast, the isoelectronic 4d based $Sr_2RhO_4$ with $Rh^{4+}$ ($4d^5$) ions with five 4d electrons has a weaker SOI (~ 0.16 eV), thus a smaller splitting between the $J_{eff}$ = 1/2 and $J_{eff}$ = 3/2 bands that are more evenly filled by the five 4d-electrons **[5, 15-18]**. The weaker SOI combined with more effectively screened Coulomb interactions between O-2p and Rh-4d electrons favors a metallic state **[15]**. Indeed, $Sr_2RhO_4$ is a paramagnetic, correlated metal **[16-18]** sharply contrasting the magnetic insulator $Sr_2IrO_4$ that orders at $T_C$ = 240 K **[6, 19-21]**. In addition, comparisons of $Sr_2RhO_4$ with another 4d-based compound, $Sr_2RuO_4$, a p-wave superconductor **[22],** reveal that the impact of the SOI strongly depends on the detailed band structure near the Fermi surface $E_F$, the Coulomb interactions and the lattice distortions **[7, 15, 23, 24]**. The $t_{2g}$ bands in $Sr_2RhO_4$ near $E_F$ are less dispersive than those in $Sr_2RuO_4$, therefore more susceptible to the SOI-induced band shifts near $E_F$ than in $Sr_2RuO_4$ despite the similar strength of the SOI in both materials **[23]**. This is in part because the $Ru^{4+}$ ($4d^4$) ion has **four** 4d electrons instead of five; Ru doping therefore adds holes to the bands.

Both $Sr_2IrO_4$ and $Sr_2RhO_4$ are not only isoelectronic but also isostructural with a crystal structure similar to that of $Sr_2RuO_4$ and $La_2CuO_4$ **[16]**. A unique and important structural feature shared by both $Sr_2IrO_4$ and $Sr_2RhO_4$ is that they crystallize in a reduced tetragonal structure with space-group $I4_1/acd$ due to a rotation of the $IrO_6$- or $RhO_6$-octahedra about the c-axis by ~12° or ~ 9.7°, respectively, resulting in a larger unit cell by $\sqrt{2} \times \sqrt{2} \times 2$ **[16, 19-21]** as compared to the undistorted cell.

That the two isostructural and isoelectronic compounds exhibit the sharply contrasting physical properties underscores the critical role SOI plays in determining the ground state of the iridate. In this work, we tune the ground state via reducing SOI by substituting $Rh^{4+}(4d^5)$ for $Ir^{4+}(5d^5)$ in $Sr_2IrO_4$, i.e., in single-crystal $Sr_2Ir_{1-x}Rh_xO_4$ ($0 \le x \le 1$). Unlike other chemical substitutions, the Rh substitution directly reduces the SOI, thus the splitting between the $J_{eff} = 1/2$ and $J_{eff} = 3/2$ bands but without obviously altering the band filling. Hence, the system remains tuned at the Mott instability and is very susceptible to disorder scattering which gives rise to localization. For comparison and contrast, we also substitute $Ru^{4+}(4d^4)$ for $Ir^{4+}(5d^5)$ in $Sr_2IrO_4$ i.e., $Sr_2Ir_{1-x}Ru_xO_4$ ($0 \le x \le 1$) where Ru not only reduces the SOI but also fills the $t_{2g}$ bands with holes, which lowers $E_F$, thus moving the system away from the Mott instability. Disorder scattering is then less relevant, and Ru doping systematically drives the system to a robust metallic state. The anticipated underlying effects of Rh and Ru doping on the $J_{eff} = 1/2$ and $J_{eff} = 3/2$ bands are schematically illustrated in **Fig.1a**. The doping profoundly alters the balance between the competing local energies, namely, the SOI is weakened, while the tetragonal CEF and the Hund's coupling $J_H$ are increased. In addition, the Rh and Ir atoms are randomly distributed over the octahedra, hindering the hopping of the d-electrons because of a mismatch of the energy levels and a mismatch of the rotation of the octahedra. The resulting disorder scattering gives rise to localized states. The combined effects produces a rich T-x phase diagram in $Sr_2Ir_{1-x}Rh_xO_4$ featuring two major effects: **(1)** Light Rh doping ($0 \leq x \leq 0.16$) effectively reduces the SOI, and prompts a simultaneous and precipitous drop in both the electrical resistivity ρ(T) and the magnetic ordering temperature $T_C$, which becomes zero at x = 0.16 from 240 K at x=0. The results indicate that the Rh concentration does provide a degree of control on the splitting between $J_{eff} = 1/2$ and $J_{eff} = 3/2$ bands. **(2)** However, heavier Rh doping ($0.24 < x < 0.85$ (±0.05)) increases localization effects in the system which fosters a return to an insulating state with anomalous magnetic behavior occurring below 0.3 K that only disappears near x=1. The magnetic state is expected to arise from the strong competition between antiferromagnetic (AFM) and ferromagnetic (FM) coupling that causes strong spin frustration. A recent optical study **[26]** on thin-film $Sr_2Ir_{1-x}Rh_xO_4$ with x up to 0.26 is qualitatively consistent with some of our results. However, the present work addresses structural and physical properties of *bulk single-crystal* $Sr_2Ir_{1-x}Rh_xO_4$ with *x ranging from 0 to 1*, which has not been reported before.

## II. Experimental

The single crystals studied were grown from off-stoichiometric quantities of $SrCl_2$, $SrCO_3$, $IrO_2$ and $RhO_2$ or $RuO_2$ using self-flux techniques. Similar technical details are described elsewhere **[6, 8-10]**. The average size of the single crystals is 1.0 x 1.0 x 0.2 $cm^3$ (see **Fig.1d**). The structures of $Sr_2Ir_{1-x}Rh_xO_4$ and $Sr_2Ir_{1-x}Ru_xO_4$ were determined using a Nonius Kappa CCD X-ray diffractometer at 90 K and 295 K. Structures were refined by full-matrix least-squares using the SHELX-97 programs **[27]**. The standard deviations of all lattice parameters and interatomic distances are smaller than 0.1%. The structure of the single crystals studied is highly ordered, as evidenced by sharp Bragg diffraction peaks ([100] and [001] directions, respectively) in **Fig.1 d**. We also checked the atom site occupancy factor on oxygen sites by running a test in which the oxygen occupancies were allowed to refine freely. They all refined to 1, suggesting that no clear oxygen vacancy was found from the refinement. Chemical compositions of the single crystals were determined using a combined unit of Hitachi/Oxford SwiftED 3000 for energy dispersive X-ray (EDX) spectroscopy. The specific heat, C(T), was measured down to 50 mK whereas the resistivity, ρ(T), and the magnetization, M(T), were measured between 1.7 K and 400 K using a Quantum Design (QD) 7T SQUID Magnetometer and a QD 14T Physical Property Measurement System, respectively.

## III. Results and Discussion

Substituting $Rh^{4+}$ for $Ir^{4+}$ results in a nearly uniform reduction in the lattice parameters a-, c-axis and the unit cell V that is shrunk by ~ 2%, as shown in **Figs. 1b** and **1c**. This behavior is expected for $Rh^{4+}$ doping because the ionic radius of $Rh^{4+}$ (0.600 Å) is smaller than that of $Ir^{4+}$ (0.625 Å). (An increase in the lattice parameters would be anticipated instead for $Rh^{3+}$ ($4d^6$) doping because of the larger ionic radius of $Rh^{3+}$, 0.670 Å.) The a-axis is compressed by 0.87% whereas the c-axis only by 0.26%, which enhances the tetragonal CEF. In addition, the Ir-O-Ir bond angle θ increases significantly near x=0.16, indicating a less distorted lattice for x>0.16. It is already established that **θ** is critical to the electronic and magnetic structure of $Sr_2IrO_4$ **[7-10, 25]**.

Rh doping effectively suppresses the magnetic transition $T_C$ from 240 K at x=0 to zero at x=0.16, as shown in **Figs. 2a, 2b** and **2c.** With increasing x, the c-axis magnetization $M_c$ becomes relatively stronger, indicating an enhanced out-of-plane spin canting **(Fig.2b)**. This change is likely due to a change in the relative strength of SOI and tetragonal CEF as an enhanced tetragonal CEF due to the increased c/a ratio (**Fig.1b**) encourages a spin configuration along the c-axis **[7]**. The Curie-Weiss temperature $\theta_{CW}$, which is estimated from a fitting of the susceptibility to a Curie-Weiss law, $\Delta\chi$, and a constant term, tracks the rapidly decreasing $T_C$ for $0 \leq x \leq 0.16$. $\theta_{CW}$ is nearly zero at x= 0.16 and then changes its sign from positive to negative as x further increases. It is remarkable that $\theta_{CW}$ is - 72 K at x=0.42 and then becomes -2 K at x=1, as shown in **Fig.2c**. If we interpret $\theta_{CW}$ as measuring the strength of the magnetic interaction, such a large value of $\theta_{CW}$ in a system without magnetic ordering even at 0.3 K implies the interplay of competing interactions and a strong suppression of magnetic ordering. This is the consequence of the Rh and Ir disorder and the changed local energies with x, such as the SOI, the non-cubic CEF and the enhanced the Hund's rule coupling, which intensify the competition between AFM and FM couplings. This could be an explanation for the disappearance of the magnetic order at x=0.16 and the appearance of the spin frustration at higher x (magnetic order below 0.3 K is observed and discussed below). It is also noted that the magnetic susceptibility $\chi(T)$ for $0.24 \leq x \leq 0.75$ below 40 K follows a power law, $\chi(T) \sim T^{-\alpha}$, with $\alpha$ increasing with x from 0.35 to 0.57, suggesting strong spin interaction among unscreened spins even at low temperatures.

The Rh doping unexpectedly generates three doping regions having distinct transport behavior, that is, *Region I: $0 \leq x \leq 0.24$, Region II: $0.24 < x < 0.85$ (±0.05), and Region III: 0.85*

*(±0.05) < x ≤ 1*. The Region III represents a metallic state occurring in a very narrow region close to x=1, i.e. $Sr_2RhO_4$, that is thoroughly discussed in Ref. 17. Here we focus on the Regions I and II, which are discussed separately below.

***Region I, 0 ≤ x ≤ 0.24:*** The electrical resistivity ρ(T) for the a- and c-axis drastically reduces by nearly six orders of magnitude at low temperatures from ~ $10^6$ Ω cm at x=0 to ~ 1 Ω cm at x = 0.07, as shown in **Fig.3a**. For 0.07< x ≤ 0.24, the a-axis resistivity $\rho_a(T)$ above 50 K exhibits metallic behavior, $d\rho_a/dT > 0$, and a largely reduced magnitude of $\rho_a(T)$ ranging from $10^{-3}$ to $10^{-1}$ Ω cm (see **Fig.3b**). $d\rho_a/dT > 0$ becomes most obvious at x = 0.11. The corresponding c-axis resistivity $\rho_c(T)$ shows a larger yet comparable magnitude, but with $d\rho_c/dT$ remaining negative, as shown in **Fig.3d**. The drastic reductions in $\rho_a(T)$ and $\rho_c(T)$ are primarily attributed to the weakened SOI because the Rh doping directly reduces the SOI and adds no holes or electrons to the bands. In addition, the vanishing magnetic state in this doping range may also help reduce the band gap because the internal magnetic field lifts the degeneracy along the edge of the AF Brillouin zone, thus facilitate the SOI to open a full gap in the presence of U **[5]**. Note that the bond angle θ, which is critical to electron hopping in general, remains essentially unchanged until x>0.16 (**Fig.1c**), therefore it should not be a predominant player for 0≤x≤0.15 (see **Fig. 2b**). Both $\rho_a(T)$ and $\rho_c(T)$ exhibit a noticeable upturn below 50 K indicating that a low-temperature metallic state is not fully realized although $\rho_a(T)$ and $\rho_c(T)$ are radically reduced by six orders of magnitude. It is also noted that $\rho_a(T)$ for x=0.24 follows variable range hopping (VRH) model, $\rho \sim \exp(1/T)^{1/2}$, below 50 K. It implies that Anderson localization comes into play at x=0.24. The persisting nonmetallic state below 50 K suggests that the band gap is not fully closed with conducting states despite the weakened SOI and the diminishing internal magnetic field. It is interesting to see that 14% of Ru doping, which not only reduces SOI but also adds holes to the bands, also fails to induce a metallic state (see Inset in **Fig.3d**).

***Region II, 0.24 < x < 0.85 (±0.05):*** If the reduction of SOI would be the only mechanism, a more metallic state would be expected with increasing x. However, both $\rho_a(T)$ and $\rho_c(T)$ increase significantly, reaching $10^5$ and $10^7$ Ω cm respectively at low temperatures for x=0.70 before dropping again to $10^{-1}$ Ω cm for x = 0.75, as shown in **Figs. 3c** and **3e**. No metallic behavior ($d\rho/dT > 0$) is observed in the entire temperature range measured for x=0.42, 0.70 and 0.75. The insulating state occurring in this region is the consequence of localization due to disorder on the Rh/Ir site in the alloy. ρ for these Rh concentrations fits the VRH $\rho \sim \exp(1/T)^{1/2}$ for 2 < T < 100 K, suggesting that weak localization due to disorder becomes significant. It sharply contrasts the well-established metallic state in $Sr_2Ir_{1-x}Ru_xO_4$ with x=0.50 (inset of **Fig.3d**). It is important to note that our oxygenated single crystals with x = 0.42, 0.70 and 0.75 exhibit essentially identical magnitude and temperature dependence of $\rho_a(T)$ and $\rho_c(T)$; this result rules out an insulating state that might be induced by oxygen deficiency. Indeed, the x-ray refinement already confirms no discernible oxygen deficiency in the single crystals studied.

The ratio of ρ(2K)/ρ(300K) for both $\rho_a(T)$ and $\rho_c(T)$ qualitatively captures the change of transport properties with Rh concentration x (see Inset in **Fig.3c**). The initial, precipitous drop in the ratio from ~ $10^6$ at x = 0 to ~ 1 near x = 0.16 signals the rapidly growing metallic state. The ratio rises again at x > 0.24, marking the return into an insulating state, before falling back for x > 0.70. The effective moment $\mu_{eff}$ essentially tracks the change of the ratio of ρ(2K)/ρ(300K). This reflects the association of localized states with the magnetic degrees of freedom (Inset in **Fig.2d**).

The temperature dependence of the specific heat C for various x is shown in **Fig.4a**. Fitting the data to $C(T) = \gamma T + \beta T^3$ for $10 < T < 50$ K yields the coefficient for the electronic contribution to C(T), γ, that systematically increases with x from 7 mJ/mole $K^2$ at x=0 to 30 mJ/mole $K^2$ at x=1. The increased γ for the insulating region $0.24 < x \leq 0.75$ is a result of the localized states in the gap, which give rise to a finite density of states (**Fig.4d)**. Remarkably, C(T)/T exhibits a pronounced peak near $T_M$ = 100 mK and 280 mK for x=0.42 and 0.70, respectively, which can be completely suppressed by a magnetic field H of 9 T (**Fig.4b**). This anomaly signals a transition to a low-T spin order from a higher-T spin frustration characterized by a frustration parameter $f = |\theta_{CW}|/T_M = |-72|/0.1 = 720$ for x=0.42, for example. In contrast, $Sr_2Ir_{1-x}Ru_xO_4$ behaves more normally (**Fig.4c**), yielding γ considerably larger than that for $Sr_2Ir_{1-x}Rh_xO_4$ (**Fig.4d**), which is consistent with the robust metallic state.

**Fig.5** shows a phase diagram for $Sr_2Ir_{1-x}Rh_xO_4$ generated based on the data presented above which summarizes the central findings of this study. The initial Rh doping effectively reduces the SOI, or the splitting between the $J_{eff}$ = 1/2 and $J_{eff}$ = 3/2 bands and alters the relative strength of the SOI and the tetragonal CEF that dictates the magnetic state, which, in turn, affects the band gap near $E_F$. In addition, the Rh doping also enhances the Hund's rule coupling that competes with the SOI, and prevents the formation of the $J_{eff}$ = 1/2 state **[5]**. It is these SOI-induced changes that account for the simultaneous, precipitate decrease in ρ(T) and $T_C$ that vanishes at x=0.16. As x increases further, the Rh/Ir disorder on the transition metal site determines the properties of the system. There is an energy level mismatch for the Rh and Ir sites that makes the hopping of the carriers between an octahedron containing a Rh atom and one with an Ir ion more difficult and also changes the orientation angles of the octahedra. The randomness of the Rh/Ir occupations gives rise to Anderson localization and an insulating state for 0.24 $<x<0.85$ (±0.05). In addition, the SOI may no longer be strong enough to support the $J_{eff} = 1/2$ insulating state and the Hund's rule coupling is enhanced (on the Rh sites), hence further strengthening the competition between AFM and FM couplings. As a result of this competition, spin frustration arises at intermediate temperatures. The occurrence of a spin ordered state below 0.3 K along with the high $\theta_{CW}$ corroborates the frustrated state. These effects diminish with disappearing disorder when x approaches 1, where the weakened SOI is comparable to other relevant energies yielding a metallic state. This point is qualitatively consistent with the recent theoretical studies for $Sr_2RhO_4$ **[15, 20, 21]**.

In contrast, there is no discernible effect due to disorder in $Sr_2Ir_{1-x}Ru_xO_4$. While for isoelectronic Rh substitution the system always remains in the proximity to the Mott condition for an insulator, each Ru atom adds one hole, giving rise to a higher density of states near $E_F$ and hence supporting a more robust metallic state in $Sr_2RuO_4$. Under these circumstances disorder in the alloy plays a less relevant role.

GC is very grateful to Drs. G. Khaliulin, X. Dai, Y. B. Kim, H.Y. Kee, and G. Jackeli for enlightening discussions. This work was supported by NSF through grants DMR-0856234 (GC) and EPS-0814194 (GC, RKK) and DMR-1056536 (RKK) and by Department of Energy through grant No. DE-FG02-98ER45707 (PS).

# *High school student on internship, Paul Lawrence Dunber High School, Lexington, KY 40513*

* *Corresponding author; cao@uky.edu*

**Captions**

**Fig.1. (a)** The schematics for the effects of Rh and Ru doping on the $J_{eff} = 1/2$ and $J_{eff} = 3/2$ bands; the Rh concentration x dependence at 90 K of **(b)** the lattice parameters a-, and c-axis (right scale), and **(c)** the unit cell volume V and the Ir-O-Ir angle θ (right scale); **(d)** some representative single-crystal $Sr_2Ir_{1-x}Rh_xO_4$ and **(e)** their Bragg diffraction peaks ([100] and [001] directions); note the highly ordered crystal structure.

**Fig.2.** The temperature dependence at $\mu_oH$=0.1 T of the magnetization **(a)** $M_a$, **(b)** $M_c$ for $0 \le x \le 0.15$; and **(c)** $M_a$ for $0.24 \le x \le 0.75$ and $\Delta\chi_a^{-1}$ (right scale) for x=0, 0.42, and 0.70; **(d)** The Rh concentration x dependence of $T_C$ and $\theta_{CW}$, and the magnetic effective moment $\mu_{eff}$ (Inset).

**Fig.3.** The temperature dependence of **(a)** the resistivity ρ for x=0 and 0.07; (**b)** the a-axis resistivity $\rho_a$ for x=0.11, 0.15 and 0.24, **(c)** $\rho_a$ for x=0.42, 0.70 and 0.75; **(d)** the c-axis resistivity $\rho_c$ for x=0.11 and 0.15, **(e)** $\rho_c$ for x=0.42, 0.70 and 0.75. Inset in (a): the ratio of ρ(2K)/ρ(300K) vs x; inset in (c): ln $\rho_a$ vs $T^{-1/2}$ , and inset in (d): $\rho_a$ vs. T for $Sr_2Ir_{1-x}Ru_xO_4$, where a robust metallic state occurs at x=0.50.

**Fig.4**. **(a)** The specific heat C(T)/T vs. $T^2$ for $Sr_2Ir_{1-x}Rh_xO_4$; **(b)** C(T)/T vs. T for 50 mK<T< 20 K at $\mu_oH$=0 for x=0.42 and 0.70, and 9 T for x=0.42; **(c)** C(T)/T vs. $T^2$ for $Sr_2Ir_{1-x}Ru_xO_4$ for comparison; **(d)** γ vs. x for $Sr_2Ir_{1-x}Rh_xO_4$ and $Sr_2Ir_{1-x}Ru_xO_4$.

**Fig.5.** The phase diagram for $Sr_2Ir_{1-x}Rh_xO_4$ generated based on the data presented above.

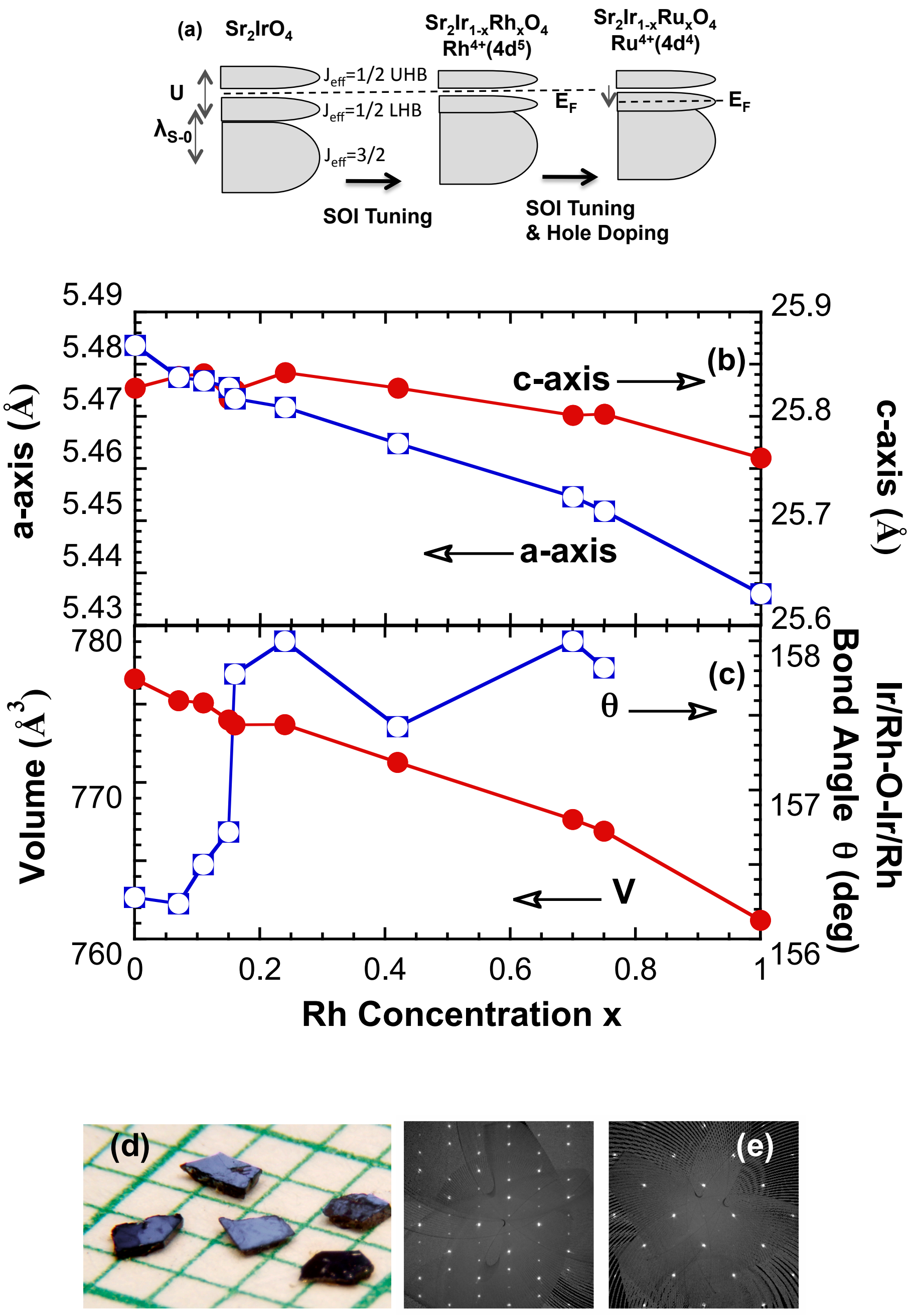

Fig.1

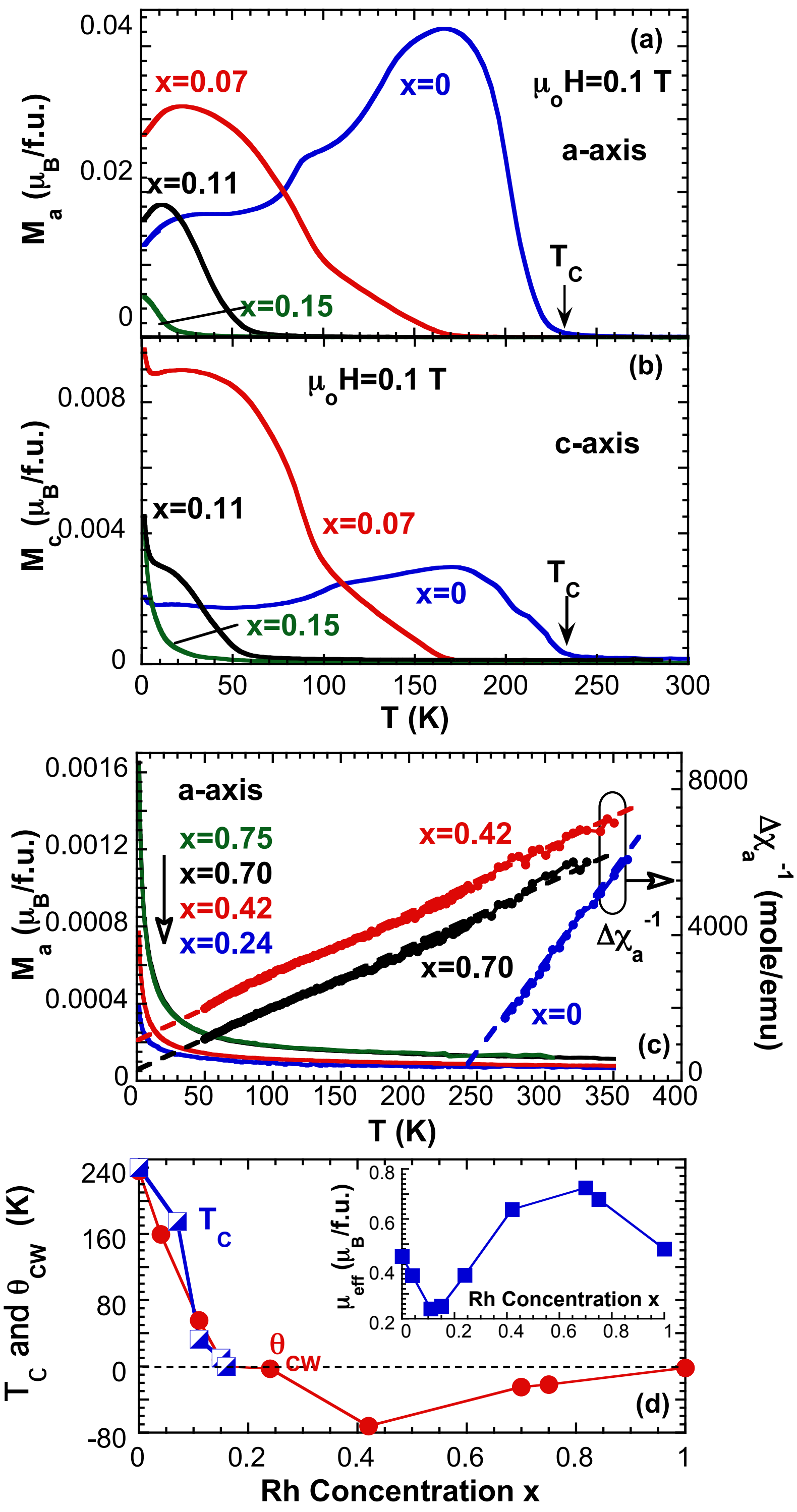
(a)
x=0.07
x=0
μoH=0.1 T
a-axis
x=0.11
Tc
x=0.15
Ma (μB/f.u.)
0.04
0.02
0
(b)
μoH=0.1 T
c-axis
x=0.11
x=0.07
x=0
Tc
x=0.15
Mc (μB/f.u.)
0.008
0.004
0
0 50 100 150 200 250 300
T (K)
a-axis
x=0.75
x=0.70
x=0.42
x=0.24
x=0.42
x=0.70
x=0
Δχa-1
(c)
Ma (μB/f.u.)
0.0016
0.0012
0.0008
0.0004
0
8000
4000
0
Δχa-1 (mole/emu)
0 50 100 150 200 250 300 350 400
T (K)
Tc and θcw (K)
240
160
80
0
-80
Tc
θcw
μeff (μB/f.u.)
Rh Concentration x
(d)
0 0.2 0.4 0.6 0.8 1
Rh Concentration x

Fig.2

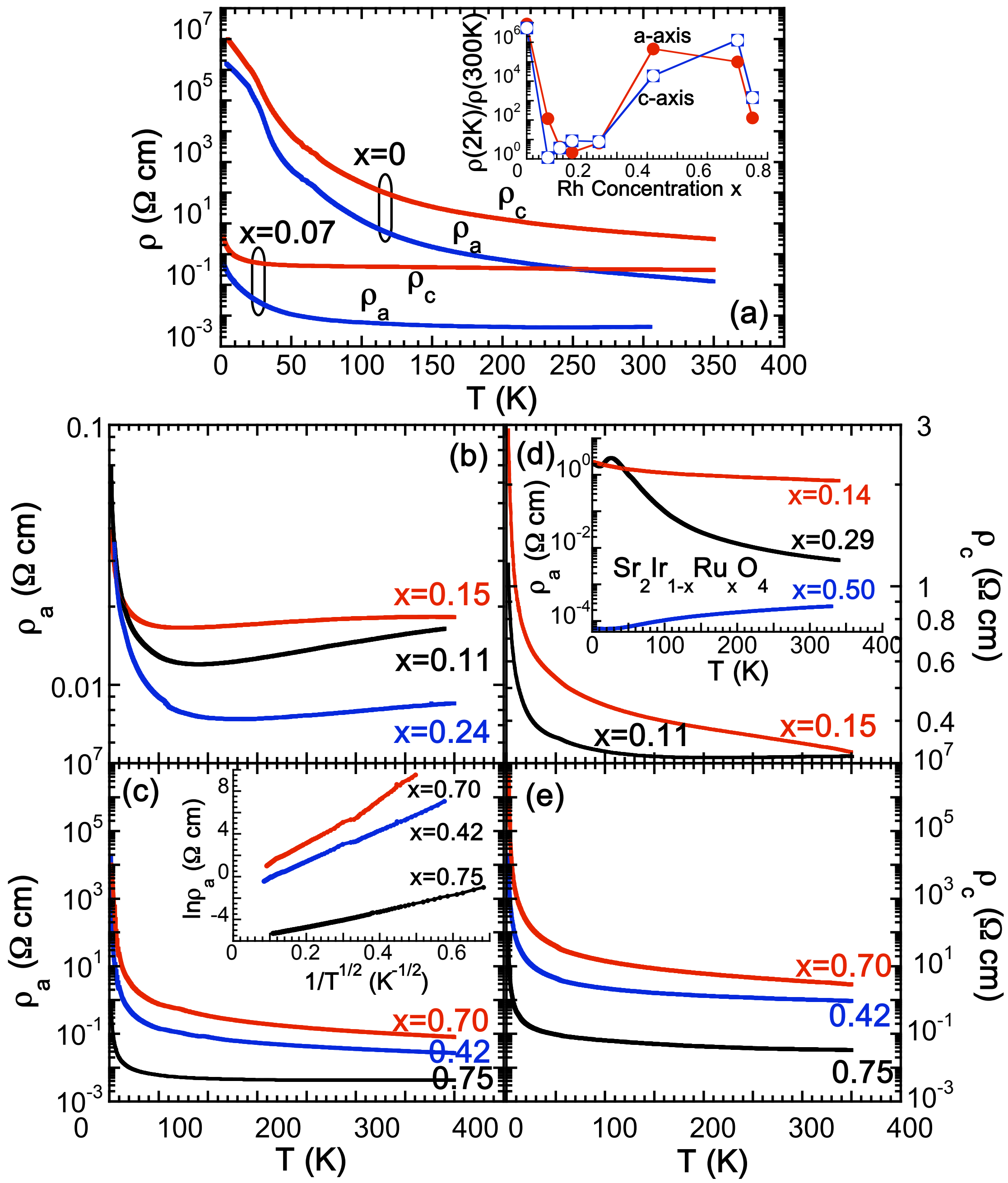

Fig. 3

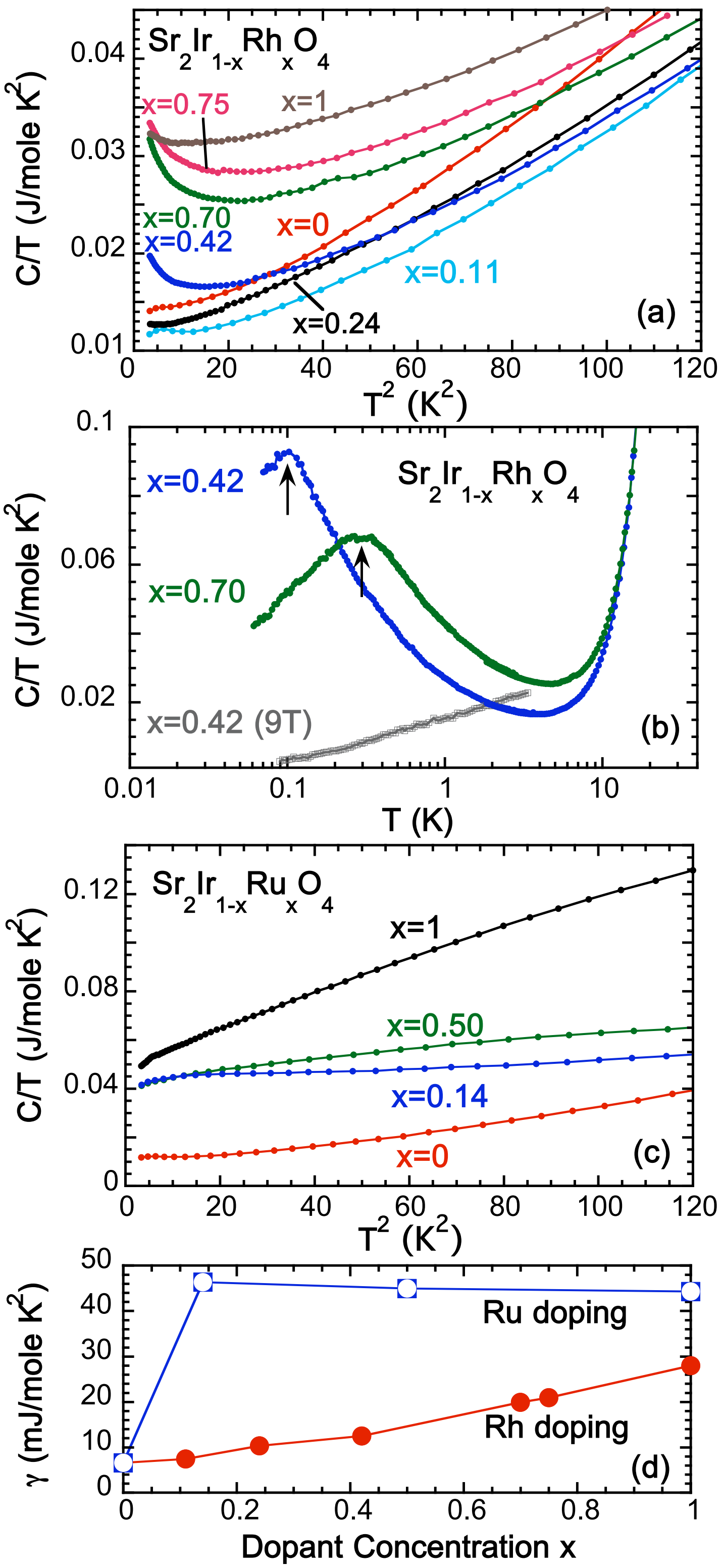

Fig.4

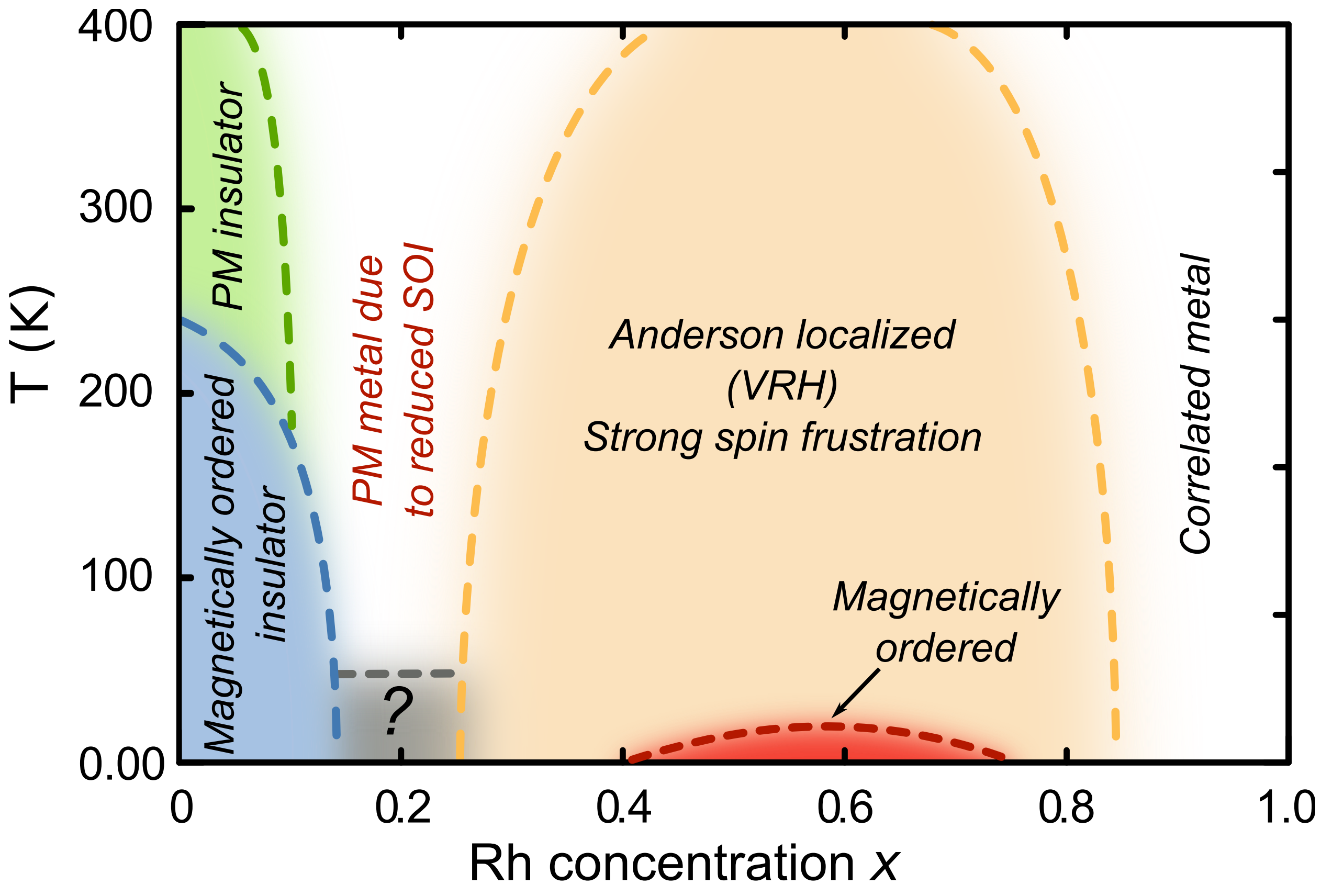

Fig.5